\documentclass[12pt]{article}%
\usepackage[T2A]{fontenc}
\usepackage[cp1251]{inputenc}
\usepackage{amsmath}
\usepackage{amsfonts}
\usepackage{amssymb}
\usepackage{graphicx}
\usepackage{cite}%
\usepackage{authblk}

\begin{document}

\title{Beamforming in multipath environment using the stable components of wave field}
\date{\ }

\author{A.L. Virovlyansky}
\affil{\footnotesize Institute of Applied Physics, Russian Academy of Science, 46 Ul'yanov Street, Nizhny Novgorod, Russia, viro@appl.sci-nnov.ru}


\maketitle

\begin{abstract}
The paper describes the beamforming procedures in an acoustic waveguide based
on representing the field on the antenna as a superposition of several stable
components formed by narrow beams of rays [A.L. Virovlyansky, J. Acoust. Soc. Am. {\bf 141}, 1180--1189 (2017)]. A modification of the matched field processing method is proposed, based on the transition from comparing the measured and calculated fields on the antenna to comparing their stable components. The modified approach becomes less sensitive to the inevitable inaccuracies of the environmental model. In the case of a pulsed source, the stable components carry signals whose arrival times can be taken as input parameters in solving the inverse problems. The use of the stable components as the initial fields on the aperture of the emitting antenna makes it possible to excite narrow continuous wave beams propagating along given ray paths.
\end{abstract}

\newpage

\section{\label{sec:1} Introduction}
The sound field at the antenna aperture in a multipath environment is a superposition of several waves coming from different directions \cite{VTrees2002}. Each wave is formed by a beam of rays hitting the aperture which we will call an eigenbeam. If the eigenbeam is so narrow that its width is small compared to the spatial scale of a weak sound speed perturbation, then, passing through the perturbation, all the rays forming this eigenbeam acquire approximately the same phase increments $\phi$. In Ref. \cite{V2017}, the field component formed by such a beam is called stable. If the wave field is excited by a tonal source, then the stable components in the perturbed and unperturbed waveguide differ by only a constant phase factor $\exp(i\phi)$. 
In the case of transient wave field the perturbation causes only an additional time delay of the stable component as a whole. This paper considers the use of vertical antennas for receiving and emitting sound waves in an underwater waveguide and proposes beamforming procedures based on the use of the stable field components.

An analog of conventional plane-wave beamforming in an inhomogeneous medium is the matched field processing (MFP) \cite{Hinich73,Bucker76}. This method is based on comparing the vector $\mathbf{v}$, whose elements are the complex amplitudes of the signals measured by the antenna elements, with the vector $\mathbf{u}$  representing the theoretical estimate of $\mathbf{v}$ calculated using the available environmental model. When solving the problem of source localization \cite{JKPS2011, Baggeroer1993} and/or reconstruction of environmental parameters \cite{Tolstoy1992}, the measured vector $\mathbf{v}$ is compared with the vectors $\mathbf{u}$ calculated for different source positions and/or different values of unknown environmental parameters. The desired estimates are given by the values of the source coordinates and medium parameters corresponding to the maximum of the scalar product of  $\mathbf{u}$ and $\mathbf{v}$. However, this approach is effective only in the case of a fairly accurate environmental model. Otherwise, that is, under conditions of uncertain environment, it may turn out that the maximum of the scalar product corresponds to values of unknown parameters that differ significantly from the true ones.

A number of approaches have been developed for working in uncertain environments. One of them is based on reducing the sensitivity of MFP through the use of a multiply constrained beamformer \cite{Schmidt1990}. This suggests that opening up the search window in one or more of these parameters would make the beamformer more tolerant of uncertainty in the other parameters \cite{Krolik1992,Baggeroer1993,Byun2020}. Another well-known approach is based on solving the problem of localization by incorporating environmental variability a priori \cite{Nolte1999,Dosso2008,Dosso2011}. The current state of research related to the use of MFP in ocean acoustics is presented in the review \cite{Sazontov2015}.

This paper discusses an alternative approach, which we call the generalized matched field processing (GMFP). Its idea is to move from comparing vectors $\mathbf{v}$ and $\mathbf{u}$ to comparing their stable components. This relaxes the accuracy requirements for the environment model. The use of GMFP is illustrated by a numerical example.

In the case of a pulsed source, a procedure has been proposed for isolating signals carried by stable components, that is, arriving at the antenna through individual eigenbeams. The arrival times of these signals can be used as input parameters in solving inverse problems.

It is also shown that the use of stable components in combination with the phase conjugation method \cite{Jackson91,Fink2001}, allows one to create the field distributions on the aperture of the radiating antenna for emitting narrow wave beams, propagating along given ray paths. 

The paper is organized as follows. The representation of the field at the aperture of the receiving antenna as a superposition of stable components is introduced in Sec. \ref{sec:stable}. A method for isolating the stable components from the total field is outlined in Sec. \ref{sec:isolation}. Section \ref{sec:MFP} discusses a generalization of the MFP method. Section \ref{sec:pulse} describes a procedure of isolating the stable components from the sound field excited by a broadband source. The use of a vertical emitting antenna for exciting a narrow wave beam propagating along a given ray path is discussed in Sec. \ref{sec:emit}. The results of the work are summarized in Sec. \ref{sec:conclusion}.

\section{\label{sec:stable} Field on the antenna as a superposition of stable components}
To illustrate the general statements discussed in this and subsequent
sections, we will use an idealized 2D model of an underwater acoustic waveguide in the deep sea with coordinates $(r,z)$, where $r$ is the distance and $z$ is the depth.  The z-axis is directed downward, the water surface is in the plane $z=0$, and the bottom in the plane $z=h$.

\begin{figure*}
	\includegraphics[width=15 cm]{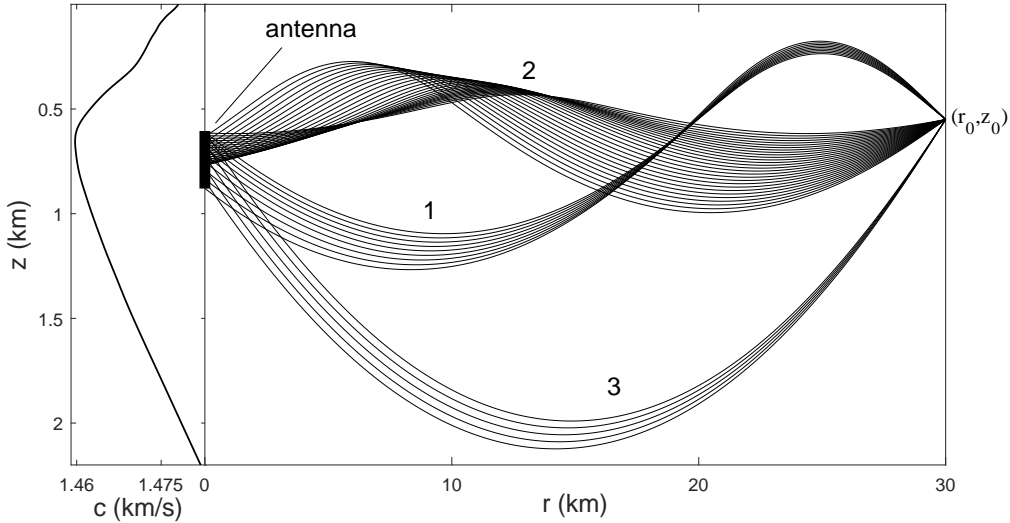}
	\caption{\label{fig:FIG1}{Left panel. Background sound speed profile $c_{b}\left(  z\right)  $. Right panel. Beams of ray (eigenbeams) hitting the antenna aperture. Next to each eigenbeam, its number is indicated.
	}}
\end{figure*}

The vertical antenna is located along the straight line $r=0$ and covers the
depth interval
$ 
z_{\text{up}}\leq z\leq z_{\text{do}}.
$
We will consider the sound speed field in the form 
$
c\left(  r,z\right)  =c_{b}\left(  z\right)  +\delta c\left(  r,z\right)  ,
$
where $c_{b}\left(  z\right)  $ is the background sound speed profile and
$\delta c\left(  r,z\right)  $ is the sound speed fluctuation. As an example,
in this work, we use a waveguide of depth h = 3 km with the profile
$c_{b}\left(  z\right)  $ shown in the left panel of Fig.~\ref{fig:FIG1}. The refractive
index is $\nu\left(  r,z\right)  =c_{0}/c\left(  r,z\right)  $, where $c_{0}$
= 1.5 km/s is the reference sound speed.

In what follows we consider the sound field excited by the source set at the
point $\left(  r_{0},z_{0}\right)  $, where $r_{0}$ = 30 km and $z_{0}$ = 0.7
km. The field is recorded by a vertical antenna with a length of $L$ = 0.25
km. The coordinates of its end points are $z_{\text{up}}$ = 0.75 km and
$z_{\text{do}}$ = 1 km. It is assumed that the bottom is strongly absorbing and therefore the contributions of the bottom reflected waves are negligible. We will consider the CW fields at the carrier frequency $f$ = 500 Hz. When simulating pulsed signals, this will be the center frequency.

The launch angles $\chi_{0}$ of rays propagating without reflections from the absorbing bottom satisfy the condition $\left\vert \chi_{0}\right\vert <\chi_{\max}$, where $\chi_{\max}$ is the critical angle \cite{JKPS2011}. It is assumed that there are $N$ eigenbeams arriving at the antenna. The
launch angles of the rays forming the $n$-th eigenbeam fill the subinterval
$\chi_{0,n}^{\prime}<\chi_{0}<\chi_{0,n}^{\prime\prime}$ of the angular
interval $\left(  -\chi_{\max},\chi_{\max}\right)  $. The subintervals
corresponding to different $n$ do not overlap. If the antenna length tends to
zero, then eigenbeams turn into ordinary eigenrays, that is, into rays
arriving at a given point.

In our example, $\chi_{\max}=12.5^{\circ}$ and there are three eigenbeams ($N$ = 3) shown in the right panel of Fig.~\ref{fig:FIG1}. Launch angles of eigenbeams 1, 2, and 3 lie in the intervals $\left(  -5.9^{\circ},-5^{\circ}\right)  $, $\left(
1.1^{\circ},4.4^{\circ}\right)  $, and $\left(  9.2^{\circ},9.7^{\circ
}\right)  $, respectively.

The rays belonging to the $n$-th eigenbeam in the unperturbed waveguide form
the component of the total field, which we denote by $u_{n}\left(  z\right)
$. The total field on the antenna is a superposition of these components%
\begin{equation}
u\left(  z\right)  =\sum_{n=1}^{N}u_{n}\left(  z\right)  . \label{u-un}%
\end{equation}
If the eigenbeam is narrow enough, then it forms a component, which in
Ref. \cite{V2017} is called stable. Let us dwell on this issue.

In the geometric optics approximation, the contribution to the total field
from the ray with the launch angle $\chi_{0}$ is $A\left(  \chi_{0}\right)
\exp\left[  ikS\left(  \chi_{0}\right)  \right]  $, where $A\left(  \chi
_{0}\right)  $ and $S\left(  \chi_{0}\right)  $ are the amplitude and eikonal
of the ray at the observation range, respectively, and 
$ k=2\pi f/c_{0} \label{k-ref} $ is the reference wavenumber \cite{JKPS2011,BL2003}. On short paths, a
variation of the ray trajectory and amplitude in the presence of a weak
perturbation $\delta c$ can be neglected \cite{BL2003,FDMWZ79}. In this case,
the influence of the perturbation is taken into account by replacing the
eikonal $S\left(  \chi_{0}\right)  $ by $S\left(  \chi_{0}\right)  +\delta
S\left(  \chi_{0}\right)  $, where%

\begin{equation}
\delta S\left(  \chi_{0}\right)  =-\frac{1}{c_{0}}\int_{\Gamma_{0}}\delta c~ds
\label{dS}%
\end{equation}
is the eikonal increment, $\Gamma_{0}$ is the unperturbed ray path, $ds$ is the arc length. In a deep ocean at frequencies of order 100
Hz this approximation is applicable at ranges up to a few hundred kilometers \cite{FDMWZ79}.

If the vertical spread of the $n$-th eigenbeam does not exceed
the vertical scale of the perturbation $\delta c$, all rays from this
eigenbeam intersect approximately the same inhomogeneities and acquire
approximately the same eikonal increments $\delta S\left(  \chi_{0}\right)  $.
Then the phase increments of these rays, $\phi\left(  \chi_{0}\right)
=k\delta S\left(  \chi_{0}\right)  $, are close to the same value, which we
denote $\phi_{n}$. If this condition is satisfied, we call the component
$u_{n}\left(  z\right)  $ stable. The influence of the perturbation is
manifested only in the multiplication of $u_{n}\left(  z\right)  $ by the
phase factor $\exp\left(  i\phi_{n}\right)  $ independent of $z$. The values
of $\left\vert \phi_{n}\right\vert $ are not necessarily small. They can
significantly exceed $\pi$.

If the eigenbeam is not narrow enough, it can be divided into several narrower eigenbeams. In the limiting case, when the antenna covers the entire cross-section of the waveguide, the
entire interval of launch angles $\left(  -\chi_{\max},\chi_{\max}\right)  $
can be divided into small subintervals so that each of them corresponds to a stable component. This situation was considered in Refs. \cite{V2018,V2019a}.

If all components $u_{n}\left(  z\right)  $ are stable, then in the presence
of the perturbation, the field $u\left(  z\right)  $ at the antenna becomes
\begin{equation}
v\left(  z\right)  =\sum_{n=1}^{N}\gamma_{n}u_{n}\left(  z\right)  ,
\label{v-un}%
\end{equation}
where $\gamma_{n}=\exp(i\phi_{n})$. Since different eigenbeams intersect
different inhomogeneities, the random values of $\phi_{n}$ for different $n$ are independent.

As a quantitative characteristic of the spread of $\phi_{n}$ corresponding to
different rays forming the $n$-th eigenbeam, we take%
\begin{equation}
\Delta\phi_{n}=k\left(  \frac{1}{\Delta\chi_{n}}\int_{\chi_{0,n}^{\prime}%
}^{\chi_{0,n}^{\prime\prime}}\left\langle q^{2}(\chi_0)%
\right\rangle d\chi_{0}\right)  ^{1/2}, \label{D-phi}%
\end{equation}
where $q(\chi_0)=\delta S\left(  \chi_{0}\right)  -\delta S\left(  \bar{\chi}_{n}\right)$, $\bar{\chi}_{n}=(\chi_{0,n}^{\prime\prime}+\chi_{0,n}^{\prime})/2$ is
the launch angle of the central ray, $\Delta\chi_{n}=\chi_{0,n}^{\prime\prime
}-\chi_{0,n}^{\prime}$, the angular brackets denote averaging over the random realizations of $\delta c$. Strictly speaking, Eq. (\ref{v-un})\emph{ }is
valid if
\begin{equation}
\Delta\phi_{n}\ll\pi. \label{strong}%
\end{equation}

In this paper, we model the perturbation $\delta c\left(  r,z\right)  $ by a
Gaussian random function with zero mean ($\left\langle \delta c\right\rangle
=0$) and the correlation function%
\[
\left\langle \delta c\left(  r,z\right)  \delta c\left(  r^{\prime}z^{\prime
}\right)  \right\rangle =\delta c_{\text{rms}}
\]
\[
\times
\exp\left(  -\frac{\pi\left(
	r-r^{\prime}\right)  ^{2}}{l_{r}^{2}}-\frac{\pi\left(  z-z^{\prime}\right)
	^{2}}{l_{z}^{2}}\right)  ,
\]
where $\delta c_{\text{rms}}$ is the rms amplitude of the sound speed
fluctuations, $l_{r}$ and $l_{z}$ are the horizontal and vertical correlation
scales, respectively. This simplest model significantly differs from more
realistic models used in ocean acoustics to describe the sound speed
fluctuations in the deep sea \cite{FDMWZ79,Colosi2016}. Nevertheless, as shown
in Ref. \cite{V2018}, it is suitable for the general analysis of stable components. In what follows we take $\delta
c_{\text{rms}}$ = 0.25 m/s, $l_{r}$ = 5 km, $l_{z}$ = 0.5 km.

Using this model for the 1st, 2nd, and 3rd eigenbeams shown in the right panel of Fig.~\ref{fig:FIG1}, we find $\Delta\phi_{1}$ = 0.73, $\Delta\phi_{2}$ = 1.83, and $\Delta\phi_{3}$ = 0.69, respectively. Thus, in our example, the condition (\ref{strong}) is not met. However, we assume that if the weaker condition
\begin{equation}
\Delta\phi_{n}<\pi\label{weak}%
\end{equation}
is satisfied, the field $v\left(  z\right)  $ can be approximately represented
as a linear combination of stable components $u_{n}\left(  z\right)  $ with
weight factors $\gamma_{n}$ different from $\exp\left(  i\phi_{n}\right)  $.

Representation of the field on the antenna as a superposition of stable
components underlies the beamforming procedures discussed below. In order to
use this representation one should calculate the functions $u_{n}\left(
z\right)  $. In free space, where there is only one eigenbeam ($N=1$) the
solution is obvious. If the antenna is located far from the
source, the eigenbeam is formed by almost parallel straight lines and
$u_{1}\left(  z\right)  $ is a fragment of a plane wave. However, in a
refractive medium, the situation becomes more complicated: a beam of rays
describes a wave whose phase front curvature depends on the range and,
generally, varies from zero to infinity. In such a medium, there are caustics
in the vicinity of which the ray approximation fails
\cite{JKPS2011,BL2003,Alonso2010}.

Since the stable components are introduced using the ray-based
representation of the wave field, we have no rigorous definition of
$u_{n}\left(  z\right)  $. Two heuristic methods for isolating stable
components from the total field $u\left(  z\right)  $ are proposed in Ref.
\cite{V2017}. Both methods come from the relationship between the ray-based
and wave-based descriptions of sound fields and give similar results. In this
paper, we will use one of these methods outlined in the next section.

\section{Isolation of field component formed by a beam of rays
	\label{sec:isolation}}

This section presents the method of isolating the stable components using  the coherent state expansion developed in quantum mechanics
\cite{Glauber2007,Klauder,Shl2001}.

\subsection{Coherent state expansion}

To describe the ray trajectories, we apply the Hamiltonian formalism
\cite{V2007a,Vbook2010}. In the scope of this formalism, the trajectory at a
range $r$ is defined by its depth $z$ and momentum $p=\nu\left(  r,z\right)
\sin\chi$, where $\chi$ is the grazing angle at the point $\left(  r,z\right)
.$The functions $p\left(  r,p_{0},z_{0}\right)  $ and $z\left(  r,p_{0}%
,z_{0}\right)  $, representing the solutions of the Hamilton ray equations
with the initial conditions $p = p_{0}$ and $z = z_{0}$ at $r = r_0$, determine the ray path in the phase space $\left(r,p,z\right)$.

In the case of a source located at a point $(r_{0},z_{0})$, all the rays start
from this point with different launch angles $\chi_{0}$ and, accordingly, with
different initial momenta and $p_{0}=\nu\left(  r_{0},z_{0}\right)  \sin\chi
_{0}$. In the phase plane (momentum $P$, depth $Z$) at the observation
distance $r$, the arrival of one ray is represented by a point. These points
form a curve representing a Lagrange manifold \cite{Alonso2010}. We will call
this curve a geometric ray line or simply a ray line. It is defined
parametrically by the equations $P=p\left(  r,p_{0},z_{0}\right)  $ and
$Z=z\left(  r,p_{0},z_{0}\right)  $ with fixed $r$ and $z_{0}$.

\begin{figure*}
	\includegraphics[width=12 cm]{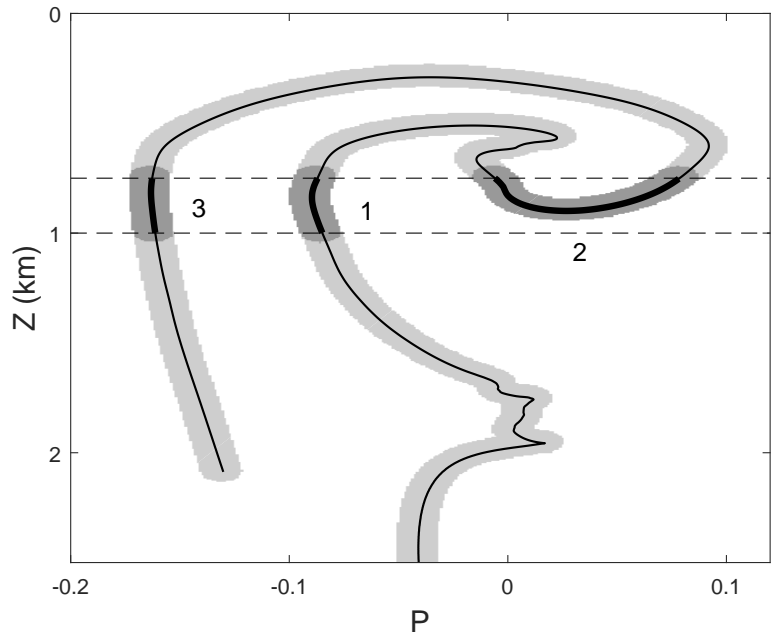}
	\caption{\label{fig:FIG2} Geometric ray line (solid curve) and fuzzy ray line (gray area) at the observation distance $r$ = 0 km. Dashed lines show the horizons of the antenna endpoints. Bold segments of the geometric ray line represent ray arrivals that hit the antenna and form the eigenbeams. Fuzzy segments corresponding to individual eigenbeams are highlighted in dark gray.
	}
\end{figure*}

In Fig.~\ref{fig:FIG2}, the solid line shows the ray line at a distance $r$ = 0 in our
example introduced in Sec. \ref{sec:stable}. The horizontal dashed lines
indicate the depths of the antenna endpoints. These straight lines
\textquotedblleft cut out\textquotedblright\ the segments of the ray line
shown by bold curves. Each segment represents the arrivals of the rays forming one of the three eigenbeams shown in Fig.~\ref{fig:FIG1}. The numbers of eigenbeams are indicated next to the corresponding segments.

To isolate the contributions of waves arriving in a neighborhood of the depth
$Z$ under grazing angles close to $\chi=\arcsin(P/\nu(r,Z))$, following
\cite{V2017,V2017a}, we use the coherent state expansion. The coherent state
associated with the point $\mu=(P,Z)$ of the phase plane is given by the
function \cite{LLquant,Klauder,Shl2001}%

\begin{equation}
Y_{\mu}\left(  z\right)  =\frac{1}{\sqrt{\Delta_{z}}}\exp\left[  ikP\left(
z-Z\right)  -\frac{\pi\left(  z-Z\right)  ^{2}}{2\Delta_{z}^{2}}\right]  ,
\label{Y-mu}%
\end{equation}
where $\Delta_{z}$ is the vertical scale.

Although the coherent states are not orthogonal, they form a complete system
of functions and an arbitrary function $u(z)$ can be represented as an
expansion \cite{Klauder,Shl2001}%
\begin{equation}
u\left(  z\right)  =\lambda^{-1}\int d\mu~a_{\mu}Y_{\mu}\left(  z\right)  ,
\label{u-Y}%
\end{equation}
where $\lambda =2\pi/k$ is the wavelength,
\begin{equation}
a_{\mu}=\int dz~u\left(  z\right)  Y_{\mu}^{\ast}\left(  z\right)  \label{a-Y}%
\end{equation}
and superscript asterisk denotes complex conjugation. The integration with
respect to $\mu$ formally goes over the entire phase plane and the integration
with respect to $z$ goes over the entire vertical axis. However, from
(\ref{Y-mu}) it is clear that the main contribution to the integral
(\ref{a-Y}) comes from the interval $Z\pm\Delta_{z}$.

The closeness of the coherent states associated with the points of the phase
plane $\mu=\left(  P,Z\right)  $ and $\mu_{1}=\left(  P_{1},Z_{1}\right)  $,
can be quantitatively characterized by their squared scalar product%
\begin{equation}
\left\vert \int dz~Y_{\mu}\left(  z\right)  Y_{\mu_{1}}^{\ast}\left(
z\right)  \right\vert ^{2}=e^{-\frac{1}{2}d\left(  \mu,\mu_{1}\right)  },
\label{Y-scalar}%
\end{equation}
where%
\begin{equation}
d\left(  \mu,\mu_{1}\right)  =\frac{\pi(P-P_{1})^{2}}{\Delta_{p}^{2}}%
+\frac{\pi\left(  Z-Z_{1}\right)  ^{2}}{\Delta_{z}^{2}}, \label{d}%
\end{equation}
$\Delta_{p}=\lambda/(2\Delta_{z})$. We
will interpret the quantity $d\left(  \mu,\mu_{1}\right)  $ as a dimensionless
distance between the points $\mu$ and $\mu_{1}$. Coherent states are close if
this distance is small compared with unity. The distance from the point $\mu$
to a curve in the phase plane (for example, to the ray line or to its segment) is the distance from $\mu$ to the nearest point of the curve.

In quantum mechanics, the wave function $Y_{\mu}(z)$ defines a state with a
minimum uncertainty \cite{LLquant,Klauder,Shl2001}. In acoustics $Y_{\mu}(z)$ can be interpreted as the cross
section of a Gaussian wave beam arriving at the grazing angle $\chi
=\arcsin\left(  P/\nu\right)  $ in a neighborhood of the depth $Z$. According
to (\ref{a-Y}) -- (\ref{d}), the coherent state amplitude, $a_{\mu}$, is
formed by contributions of waves arriving in the depth interval $Z\pm
\Delta_{z}/2$ under grazing angles corresponding to the momentum interval
$P\pm\Delta_{p}/2$.

Equations (\ref{a-Y}) -- (\ref{d}) suggest that the distribution of the squared coherent state amplitude $\left\vert a_{\mu}\right\vert ^{2}$ in the phase
plane is localized mainly in the region, formed by points located at distances%

\begin{equation}
d<1 \label{d-d0}%
\end{equation}
from the ray line. We will call this region \textbf{a fuzzy ray line} and
denote it by the symbol $\sigma$.

In Fig.~\ref{fig:FIG2}, the area occupied by the fuzzy ray line is highlighted in gray.
Here the coherent state expansion is performed with $\Delta_{z}$ = 90 m. With
this choice of $\Delta_{z}$, the area of the fuzzy ray line, and hence its
average width, in our example takes the minimum value.

\subsection{Synthesis of a stable field component from coherent states}

Consider a segment of the geometrical ray line presenting the arrivals of the
rays forming the $n$-th eigenbeam. It is natural to assume that the
contribution of these rays to the total field is represented by a
superposition of coherent states associated with points of the $P-Z$ plane
located at distances $d<1$ from the segment. We will call the area occupied by these points a fuzzy segment and denote it by $\sigma_{n}$. In Fig.~\ref{fig:FIG2}, the fuzzy segments are highlighted in dark gray.

According to this assumption, the contribution of the $n$-th eigenbeam to the
total field is%
\begin{equation}
u_{n}\left(  z\right)  =\lambda^{-1}\int_{\sigma_{n}}d\mu~a_{\mu}Y_{\mu
}\left(  z\right)  . \label{un}%
\end{equation}
Using Eq. (\ref{a-Y}), this expression can be rewritten as%

\begin{equation}
u_{n}\left(  z\right)  =\int dz^{\prime}~\Xi_{n}\left(  z,z^{\prime}\right)
u\left(  z^{\prime}\right)  , \label{Un-Q}%
\end{equation}
where%
\begin{equation}
\Xi_{n}\left(  z,z^{\prime}\right)  =\lambda^{-1}\int_{\sigma_{n}}d\mu~Y_{\mu
}\left(  z\right)  Y_{\mu}^{\ast}\left(  z^{\prime}\right)  . \label{Q-Y}%
\end{equation}
Equation (\ref{Un-Q}) explicitly defines the procedure for isolating the
desired component $u_{n}\left(  z\right)  $ from the total field $u\left(
z\right)  $ calculated or measured using a sufficiently long antenna. The
components $u_{n}\left(  z\right)  $ found in this way are determined in the
entire cross-section of the waveguide. 

Note that locally horizontal portions of the ray line correspond to caustics \cite{Alonso2010}. In Fig. 2 it is seen that such a portion is present on the bold segment formed by the rays forming the 2nd eigenbeam. In the right panel of Fig. 1 we see that the antenna crosses a caustic formed by rays from this eigenbeam. Therefore, the field component formed by this eigenbeam cannot be described in the geometrical optics approximation. However, for the applicability of the method outlined here, it is not an obstacle. The total field $u(z)$ at the observation range present in Eq. (\ref{Un-Q}) should be obtained by a full-wave calculation. In this paper, we apply the method of wide-angle parabolic equation \cite{JKPS2011}. The ray tracing is used only to find the fuzzy segment $\sigma_n$.

\begin{figure*}
	\includegraphics[width=12 cm]{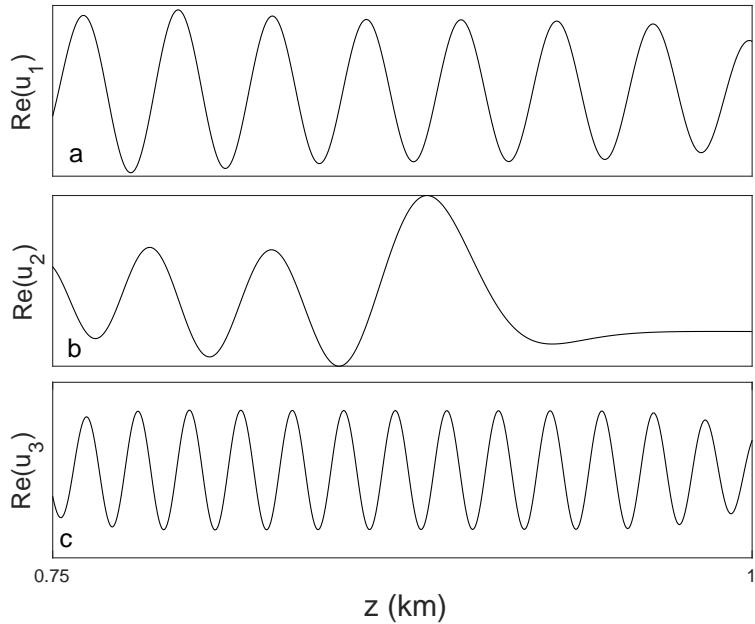}
	\caption{\label{fig:FIG3} The real parts of the functions $u_{1}\left(  z\right)  $ (a), $u_{2}\left(z\right)  $ (b), and $u_{3}\left(  z\right)  $ (c) on the antenna aperture.
	}
\end{figure*}

Figure~\ref{fig:FIG3} shows the real parts of the
field components $u_{1}\left(  z\right)  $, $u_{2}\left(  z\right)  $, and
$u_{3}\left(  z\right)  $ at the antenna aperture considered in our example.

\section{Generalized MFP \label{sec:MFP}}

The classical MFP method \cite{Hinich73,Bucker76,JKPS2011} is based on the
assumption that the available environmental model is accurate enough to
correctly calculate the field on the antenna for the known source position.
In this section, a generalized version of this method is introduced, based on
the weaker assumption that the field on the
antenna can be represented as Eq. (\ref{v-un}), that is, as the sum of the known
stable component $u_{n}\left(  z\right)  $ with unknown weights $\gamma_{n}$.

\subsection{Projection of the measured field onto stable components}

We assume that the antenna is an array consisting of a large number of
elements densely filling the aperture (e.g., ten elements per wavelength). As
in the Introduction, we will represent the calculated and measured fields by
the vectors $\mathbf{u}$ and $\mathbf{v}$, respectively, whose elements are
the complex amplitudes of the signals at the antenna elements. The proximity
of vectors $\mathbf{u}$ and $\mathbf{v}$ is quantitatively characterized by
their normalized squared scalar product%
\begin{equation}
K_{0}=\frac{\left\vert \mathbf{u}^{+}\mathbf{v}\right\vert ^{2}}{\left\vert
	\mathbf{u}\right\vert ^{2}\left\vert \mathbf{v}\right\vert ^{2}}%
=\frac{\mathbf{v}^{+}\mathbf{P}_{0}\mathbf{v}}{\left\vert \mathbf{v}%
	\right\vert ^{2}},\label{K0}%
\end{equation}
where
\begin{equation}
\mathbf{P}_{0}\mathbf{=}\frac{\mathbf{uu}^{+}}{\left\vert \mathbf{u}%
	\right\vert ^{2}}\label{P0}%
\end{equation}
is the projection matrix. The coefficient (\ref{K0}) is proportional to the
square of the projection of the measured vector $\mathbf{v}$ onto the
calculated vector $\mathbf{u}$.

If $\mathbf{u}$ is considered as a function of unknown parameters $\theta$
that specify the coordinates of the source and/or parameters of the medium,
then the similarity coefficient $K_{0}$ is also a function of $\theta$. The
maximum of this function corresponds to the actual source position and/or the
correct value of the medium parameters. This can be used in solving inverse
problems. However, as indicated above, this is true only when the model of the medium is sufficiently accurate.

Our idea is to relax the requirements for the accuracy of the environmental
model by using the field representation (\ref{v-un}). In matrix notation, this expression takes the form
\begin{equation}
\mathbf{v=}\sum_{n=1}^{N}\gamma_{n}\mathbf{u}_{n}, \label{v-un1}%
\end{equation}
where $\mathbf{u}_{n}$ are vectors representing the stable field components at the antenna aperture. Such a representation of the measured vector $\mathbf{v}$ adequately describes the situation arising from multipath sound propagation.

Note that $\mathbf{u}$ and $\mathbf{v}$ are elements of the vector space
$\Omega$ formed by vectors of size $N_{a}\times1$, where $N_{a}$ is the number
of antenna elements. The subspace of $\Omega$, whose basis is given by the
vectors $\mathbf{u}_{n}$, $n=1,\ldots,N$, we denote by $\Omega^{\prime}$. Let
us introduce the matrix $\mathbf{W}=[\mathbf{u}_{1},\ldots,\mathbf{u}_{N}]$,
whose columns are the vectors $\mathbf{u}_{n}$, and use its singular value decomposition
\cite{Golub}%
\[
\mathbf{W=}\sum_{n=1}^{N}\alpha_{n}\pmb{\xi}_{n}\pmb{\eta}_{n}^{+},
\]
where $\alpha_{n}$ are the singular numbers, $\pmb{\xi}_{n}$ and
$\pmb{\eta}_{n}$ are the singular vectors. The projection matrix
\begin{equation}
\mathbf{P=}\sum_{n=1}^{N}\pmb{\xi}_{n}\pmb{\xi}_{n}^{+} \label{P}%
\end{equation}
projects any vector from $\Omega$ onto $\Omega^{\prime}$. The new similarity
coefficient, characterizing the proximity of $\mathbf{u}$ and $\mathbf{v}$, we
define as
\begin{equation}
K=\frac{\mathbf{v}^{+}\mathbf{Pv}}{\mathbf{v}^{+}\mathbf{v}}. \label{K}%
\end{equation}
It differs from (\ref{K0}) by replacing the projection matrix $\mathbf{P}_{0}$
by $\mathbf{P}$.

If the vectors $\mathbf{u}$ and $\mathbf{v}$ correspond to the same source
position and $\mathbf{v}$ can be represented in the form (\ref{v-un1}), then
$\mathbf{Pv}=\mathbf{v}$ and $K=1$. The value of $K_{0}$ averaged over the
ensemble of sound speed fluctuations can be easily estimated under the
assumption that for $n\neq m$ the vectors $\mathbf{u}_{n}$ and $\mathbf{u}%
_{m}$ are almost orthogonal, that is,%
\begin{equation}
\left\vert \mathbf{u}_{n}^{+}\mathbf{u}_{m}\right\vert \ll\left\vert
\mathbf{u}_{n}\right\vert \left\vert \mathbf{u}_{m}\right\vert . \label{unm}%
\end{equation}
This situation is typical of an antenna whose length $L$ is large compared to
the wavelength $\lambda$. If the vectors $\mathbf{u}_{n}$ represent the
arrivals of quasi-plane waves, then (\ref{unm}) is satisfied under the
condition $\lambda/L\ll\chi_{\max}$. We also assume that the phase increments
$\phi_{n}$, $n=1,\ldots,N$, are independent random variables uniformly
distributed in the interval $\left(  0,2\pi\right)  $. Then%
\[
\left\langle K_{0}\right\rangle =\frac{\sum_{n=1}^{N}\left\vert \mathbf{u}%
	_{n}\right\vert ^{4}}{\left(  \sum_{n=1}^{N}\left\vert \mathbf{u}%
	_{n}\right\vert ^{2}\right)  ^{2}}.
\]
It follows that $\left\langle K_{0}\right\rangle $ is always in the range from $1/N$ (if all $\left\vert \mathbf{u}_{n}\right\vert $ are equal) to 1.
For individual realizations of perturbation, the value of $K_{0}$ can be less
than $1/N$.

Note that if the condition (\ref{unm}) is satisfied and $\pmb{\xi}_{n}\simeq
\mathbf{u}_{n}/\left\vert \mathbf{u}_{n}\right\vert $, then
\[
\mathbf{v}^{+}\mathbf{Pv}\simeq\sum_{n=1}^{N}\left\vert \mathbf{u}_{n}%
^{+}\mathbf{v}_{n}\right\vert ^{2},
\]
where $\mathbf{v}_{n}=\gamma_{n}\mathbf{u}_{n}$. This relation allows us to
interpret the transition from $K_{0}$ to $K$, as the transition from comparing
the measured and calculated fields to comparing their stable components.

\begin{figure*}
	\includegraphics[width=12 cm]{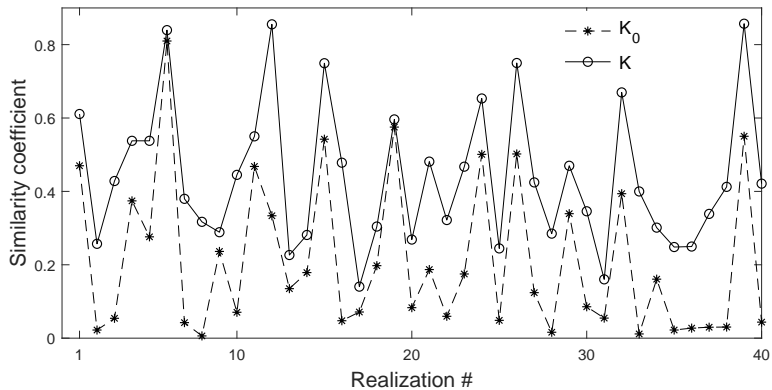}
	\caption{\label{fig:FIG4} Similarity coefficients $K_{0}$ and $K$ calculated for 40 realizations of random perturbation $\delta c\left(  r,z\right)  $. Both coefficients are found for the perturbed and unperturbed fields corresponding to the same source position $\left(  r_{0},z_{0}\right)$.
	}
\end{figure*}

Let us turn to the example described in Sec. \ref{sec:stable}. Figure~\ref{fig:FIG4}  shows the 
results of calculating the coefficients $K$ (circles) and $K_{0}$ (asterisks)
for 40 realizations of the perturbation $\delta c$. The calculations were
performed for the positions of the source and antenna shown in Fig.~\ref{fig:FIG1}. The
fact that the coefficient $K$ in Fig.~\ref{fig:FIG4}  is less than unity indicates that the
field representation (\ref{v-un1}) is not entirely accurate and the vector
$\mathbf{v}$ is only approximately described by a superposition of stable
components. Nevertheless, for each realization of $\delta c$, the value of $K$
exceeds $K_{0}$. This is consistent with our expectation that the coefficient
$K$ is less sensitive to sound speed fluctuations than $K_{0}$.

The similarity coefficient $K$ can be used to solve the same inverse problems
as $K_{0}$. The MFP method, modified by replacing $K_{0}$ with $K$, we will
call generalized matched field processing (GMFP). In the absence of multipath,
$N$ = 1 and GMFP reduces to MFP. It is natural to expect that the GMFP method
is more robust and less sensitive to inaccuracies in the environmental model.
In the next section, this will be demonstrated by a numerical example.

\subsection{Source localization}

Consider the use of MFT and GMFT to estimate the source coordinates in our
waveguide model. It is assumed that the antenna receives the signals of the tonal source placed at the point $\left(  r_{0},z_{0}\right)  $ indicated in Fig. 1. The vector $\mathbf{v}$, calculated for a realization of $\delta c$, simulates the measured field. It is compared with the fields on the antenna in the unperturbed waveguide ($\delta c=0$) excited by sources located at different points $\left(  r_{s},z_{s}\right)$. The vectors of signal
amplitudes $\mathbf{u}$, stable components $\mathbf{u}_{n}$, and matrices
$\mathbf{P}_{0}$ and $\mathbf{P}$ were computed for a set of source positions
$\left(  r_{s},z_{s}\right)  $. Then, using Eqs. (\ref{K0}) and (\ref{K}),
similarity coefficients $K_{0}$ and $K$ were found for each position. The
arguments of the uncertainty functions $K_{0}\left(  r_{s},z_{s}\right)  $ and
$K\left(  r_{s},z_{s}\right)  $, corresponding to the main maxima of these
functions, are taken as the estimates of the actual source coordinates.

\begin{figure*}
	\includegraphics[width=12 cm]{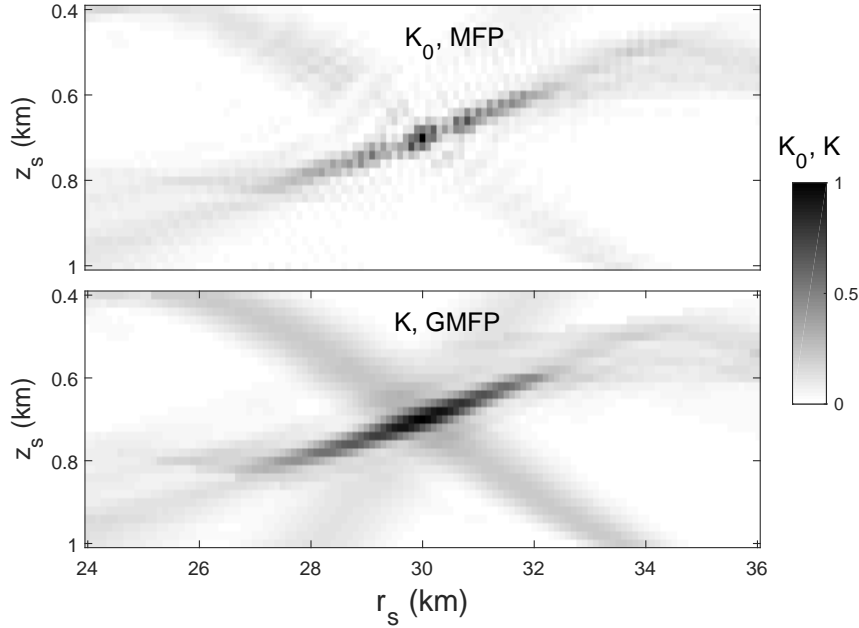}
	\caption{\label{fig:FIG5} The uncertainty functions $K_{0}\left(  r_{s},z_{s}\right)  $ (upper panel)
		and $K\left(  r_{s},z_{s}\right)  $ (lower panel) in the unperturbed waveguide ($\delta c=0$).
	}
\end{figure*}

Figure~\ref{fig:FIG5} shows the uncertainty functions in the unperturbed waveguide ($\delta c=0$). As it
should be, both functions take maximum values at the point with
coordinates $z_{s}=z_{0}=0.7$ km and $r_{s}=r_{0}=30$ km. The maximum of the
function $K(r_{s},z_{s})$ is so sharp that it is hardly distinguishable in the
figure. The function $K(r_{s},z_{s})  $ has a much wider maximum.
This indicates a weak sensitivity of the coefficient $K$ not only to the sound speed variations but also to the variations in the source position.

\begin{figure*}
	\includegraphics[width=12 cm]{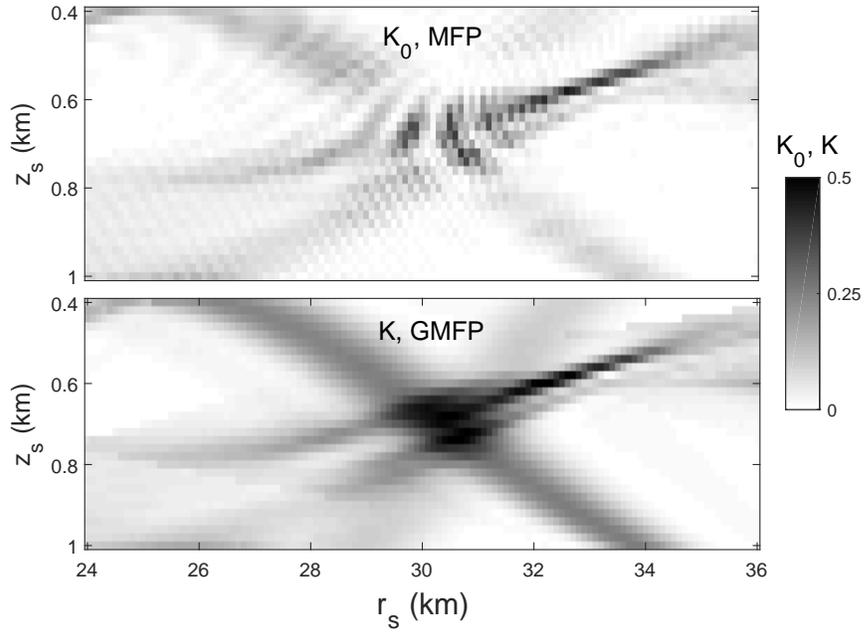}
	\caption{\label{fig:FIG6} The same as in Fig. 5, but in the presence of a realization of $\delta c\left(r,z\right)  $.
	}
\end{figure*}

Similar uncertainty functions calculated for $\delta c \neq 0$, are shown in Fig.~\ref{fig:FIG6}. In the presence of fluctuations, the main maximum of $K_ {0} (r_{s},z_{s}) $ splits into many small local maxima. The function $K(r_{s},z_{s})$ remains smooth and takes the largest values (exceeding the values of $K_{0}(r_s,z_s)$) in the vicinity of the point $(r_0, z_0)$.

\begin{figure*}
	\includegraphics[width=12 cm]{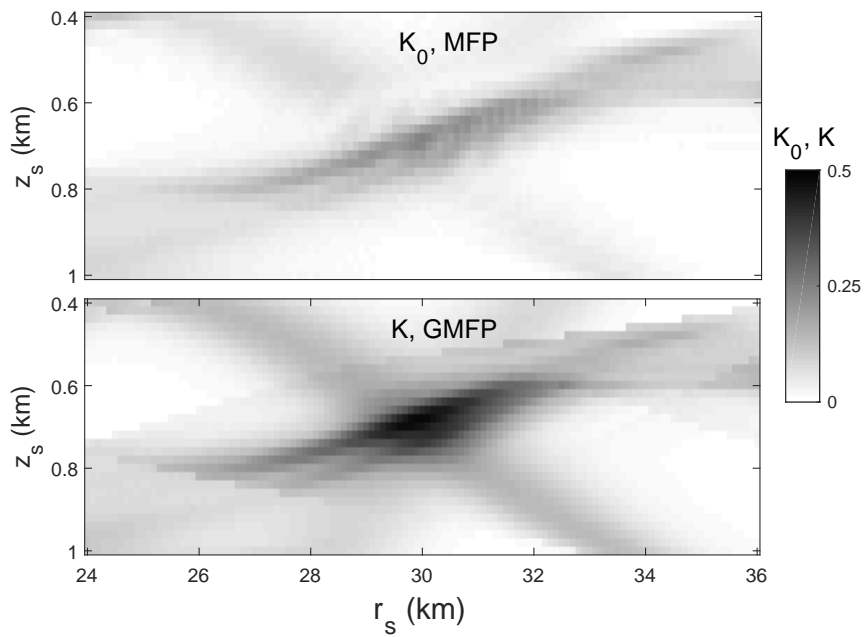}
	\caption{\label{fig:FIG7} The same as in Figs. 5 and 6, but the uncertainty functions are averaged over 40 realizations of $\delta c$.
	}
\end{figure*}

The uncertainty functions $K_{0}\left(  r_{s},z_{s}\right)  $ and $K\left(
r_{s},z_{s}\right)  $ were calculated for 40 realizations of the perturbation
$\delta c$. The functions averaged over all the realizations are shown in Fig.~\ref{fig:FIG7}. Both averaged functions have wide maxima centered approximately at the
point $\left(  z_{0},r_{0}\right)  $. However, the values {}{}of $\left\langle
K\right\rangle $ in the vicinity of this point are two times
higher than the values {}{}of $\left\langle K_{0}\right\rangle $. This is
consistent with the data presented in Fig.~\ref{fig:FIG4}, which shows the values of $K_{0}$ and $K$ at the point $\left(  r_{0},z_{0}\right)  $ for the same 40 realizations of $\delta c$. In this paper, we do not take into account the external noise. However, the results presented in Fig.~\ref{fig:FIG7} suggest that the transition from MFP to GMFP can enhance the signal-to-noise ratio at the beamformer output.

\section{Pulsed source \label{sec:pulse}}

Let us proceed to the analysis of the field excited by a pulsed point source
and consider the isolation of sound pulses arriving at the antenna via
individual eigenbeams. In this case, the fields on the antenna in a perturbed and unperturbed waveguide can be represented as%

\begin{equation}
V(z,t)  =\int df\;v(z,f)  s(f)
e^{-2\pi ift} \label{Vzt}%
\end{equation}
and%
\begin{equation}
U(z,t)  =\int df\;u(z,f)  s(f)
e^{-2\pi ift}, \label{Uzt}%
\end{equation}
respectively, where $t$ is the time, $s\left(  f\right)  $ is the spectrum of
the emitted pulse. Here we explicitly indicate the argument $f$ of functions
$v\left(  z,f\right)  $ and $u\left(  z,f\right)  $, which is omitted in the
other sections of this paper. Since the ray trajectories are frequency
independent, we will assume that the eigenbeams are the same at all frequencies.

In principle, the conditions (\ref{strong}) or (\ref{weak}) may not
be satisfied for all the frequencies from the source bandwidth. Therefore, at some frequencies, it may be necessary to split eigenbeams into narrower ones, as
indicated in Sec. \ref{sec:stable}. We do not consider such a situation,
assuming that the bandwidth is not very large.

Functions $u_{n}\left(  z,f\right)  $ representing the contributions of individual eigenbeams can be found at any frequency $f$ using Eq. (\ref{Un-Q}). To isolate the contribution of the $n$-th eigenbeam to the
total pulsed field, we will proceed as follows. Assuming that
functions $u_{n}\left(  z,f\right)  $ with different $n$ are almost
orthogonal, that is, if the condition (\ref{unm}) is satisfied at each
frequency, we calculate the projections of the field $v\left(  z,f\right)  $
on the function $u_{n}\left(  z\right)  $%
\begin{equation}
g_{n}\left(  f\right)  =\frac{\int_{z_{\text{up}}}^{z_{\text{do}}}%
	dz~u_{n}^{\ast}\left(  z,f\right)  v\left(  z,f\right)  }{\left(
	\int_{z_{\text{up}}}^{z_{\text{do}}}dz~\left\vert u_{n}\left(  z,f\right)
	\right\vert ^{2}\right)  ^{1/2}}. \label{gn}%
\end{equation}
We define the contribution of the $n$-th eigenbeam,  as%
\begin{equation}
V_{n}\left(  z,t\right)  =\int df~g_{n}\left(  f\right)  u_{n}\left(
z,f\right)  e^{-2\pi ift}. \label{Vn}%
\end{equation}

\begin{figure*}
	\includegraphics[width=12 cm]{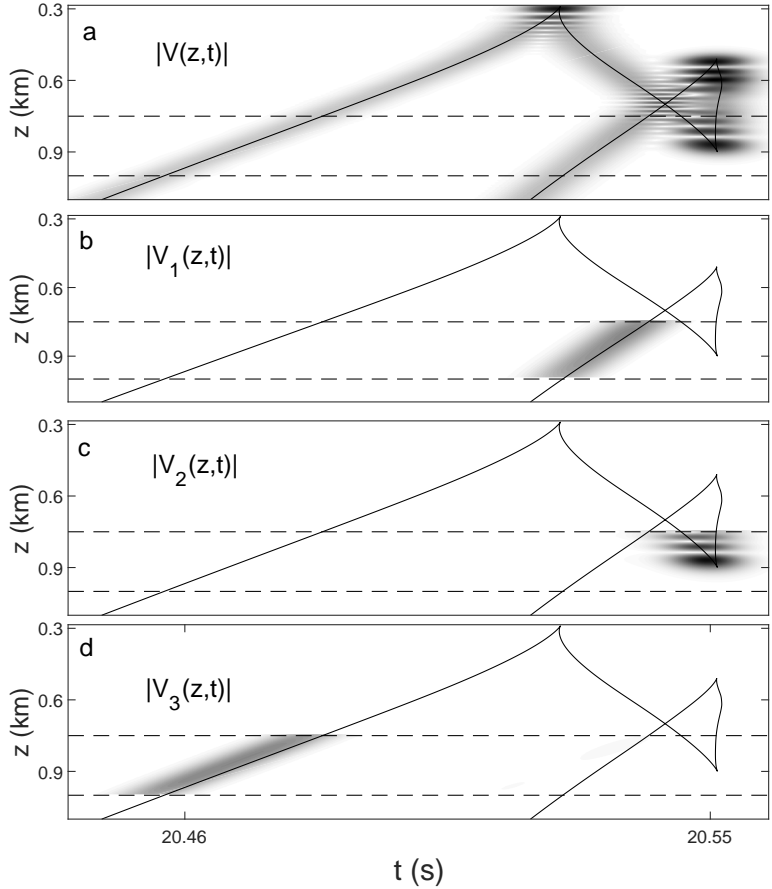}
	\caption{\label{fig:FIG8} Amplitudes of the received signals in the plane 'arrival time -- depth', $\left(t,z\right)$. The solid line, the same on all panels, represents the timefront, that is, the distribution of ray arrivals in the $\left(t,z\right)  $ plane. Dashed lines show the horizons of the antenna endpoints. (a) The total field at the observation distance $r$ = 0 in the unperturbed waveguide. (b,c,d) Pulsed field components formed by the 1-st, 2-nd, and 3-rd eigenbeams, respectively, in the presence of a realization of $\delta c(r,z)$.
	}
\end{figure*}

The results of applying this procedure in our example are presented
in Fig.~\ref{fig:FIG8}. The calculations were performed for an emitted signal with the spectrum
\[
s\left(  f\right)  =\exp\left[  -\pi\tau^{2}\left(  f-f_{0}\right)
^{2}\right]  ,
\]
where $f_{0}$ = 500 Hz, $\tau$ = 0.01 s. Figure 8a shows the distribution of the total field amplitude in the unperturbed waveguide ($\delta c=0$) in the plane "time - depth" $(t,z)$. Figures 8b, c, and d present similar distributions of the field components formed by the 1st, 2nd, and 3rd eigenbeam, respectively, in the presence of perturbation $\delta c$. The solid line, the same on all panels, shows the so-called timefront, that is, the distribution of ray arrivals in the $\left(  t,z\right)  $ plane at the observation distance $r$ = 0. Dashed lines show the depths of the antenna endpoints. The timefront segments between the dashed lines depict the arrivals of the rays belonging to the eigenbeams. Thus, we see that the described procedure allows one to isolate the field components coming through different eigenbeams. 

The pulse
\[
G_{n}\left(  t\right)  =\int df~g_{n}\left(  f\right)  e^{-2\pi ift}%
\]
can be interpreted as the total signal arriving at the antenna via the $n$-th
eigenbeam. Note the interesting fact that in the \textbf{unperturbed}
waveguide, the arrival times of all pulses $G_{n}\left(  t\right)  $ are the
same. Indeed, if $v(z,f)=u\left(  z,f\right)  $ and the condition (\ref{unm})
is satisfied at all frequencies, then%
\[
g_{n}\left(  f\right)  \simeq\left(  \int_{z_{\text{up}}}^{z_{\text{do}}%
}dz~\left\vert u_{n}\left(  z,f\right)  \right\vert ^{2}\right)  ^{1/2}%
\]
and all $G_{n}\left(  t\right)  $ take maximum values {}{}at the same moment
$t$ = 0. If the fields $U(z,t)$ and $V(z,t)$ are excited at different times $t_{u}$ and $t_{v}$, respectively, then all pulses
$G_{n}\left(  t\right)  $ will take maximum values at time $t=t_{v}-t_{u}$.

\begin{figure*}
	\includegraphics[width=12 cm]{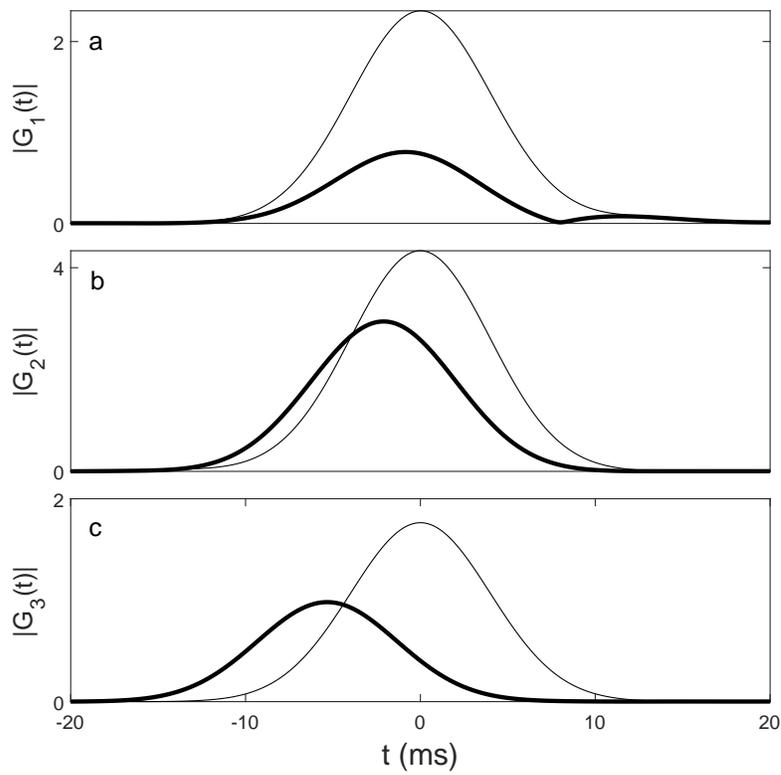}
	\caption{\label{fig:FIG9} Sound pulses arriving at the antenna via the 1-st (a), 2-nd (b), and 3-rd (c) eigenbeams in an unperturbed waveguide (thin solid lines) and in the presence of a realization of $\delta c\left(  r,z\right)  $ (thick solid lines).
	}
\end{figure*}

The pulses $G_{n}\left(  t\right)  $ calculated in the unperturbed (thin
lines) and perturbed (thick lines) waveguide are shown in Fig. 9. Figure 9a,
b, and c show the arrivals of pulses representing the contributions of the
1st, 2nd, and 3rd eigenbeams, respectively. In accordance with the above, in
an unperturbed waveguide, all pulses $G_{n}\left(  t\right)  $ arrive at the
same time. In the presence of the sound speed perturbation $\delta c$, the
arrival time of the $n$-th pulse changes by $\delta t_{n}=\delta S_{n}/c_{0}$,
where $\delta S_{n}$ is the eikonal increment, which is approximately the same
for all rays belonging to the eigenbeam. These delays are approximately equal
to those delays in the ray travel times, which are used as input parameters
in the ocean acoustic tomography \cite{MW79,Howe1987}.

\section{Emitting antenna \label{sec:emit}}

So far, we have been considering a receiving antenna. Now we turn to the
emitting antenna and show how to use a stable component for exciting  a narrow continuous wave beam propagating along a ray path.

We will apply the well-known method of phase conjugation, which is used to
focus the field at a given point \cite{Jackson91,Fink2001}. Let us denote by
$u_{0}\left(  z\right)  $ the complex amplitude of the signal emitted by the antenna element at a depth $z$. In order to focus the
field at a given point $\left(  r_{0},z_{0}\right)  $, one should take
$u_{0}\left(  z\right)  =u^{\ast}\left(  z\right)  $, where $u\left(
z\right)  $ is the field at the antenna aperture received from the source
placed at $\left(  r_{0},z_{0}\right)  $.

\begin{figure*}
	\includegraphics[width=12 cm]{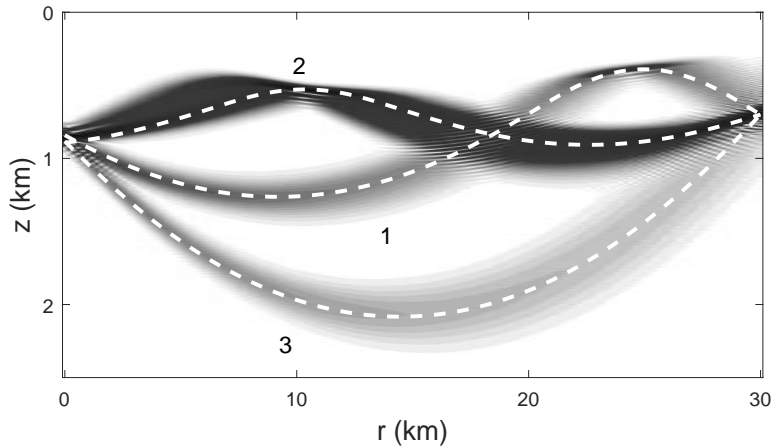}
	\caption{\label{fig:FIG10} Sound field focused at the point ($r_{0},z_{0}$) using the phase conjugation. Dashed lines show the central rays of the eigenbeams. Next to each line, the number of the corresponding eigenbeam is indicated.
	}
\end{figure*}

Consider an example of such focusing. The antenna shown in the right panel of
Fig. 1, now will be considered as the emitting one, and the source position
$\left(  r_{0},z_{0}\right)  $ will be considered as the focus point. Figure
10 presents the sound field excited in the unperturbed waveguide by this
antenna with the initial field $u_{0}\left(  z\right)  =u^{\ast}\left(
z\right)  B\left(  z\right)  $, where $B\left(  z\right)  $ is the weight
factor that ensures smooth decay of the sound field to the antenna endpoints.
We use $B\left(  z\right)  =\exp\left(  -4\pi\left(  z-z_{c}\right)
^{2}/L^{2}\right)  $, where $z_{c}=\left(  z_{\text{do}}+z_{\text{up}}\right)
/2$ is the depth of the antenna center. The excited field represents a
superposition of three wave beams propagating along the eigenbeams. The dashed
lines show the trajectories of the central rays of the eigenbeams from Fig. 1.

\begin{figure*}
	\includegraphics[width=12 cm]{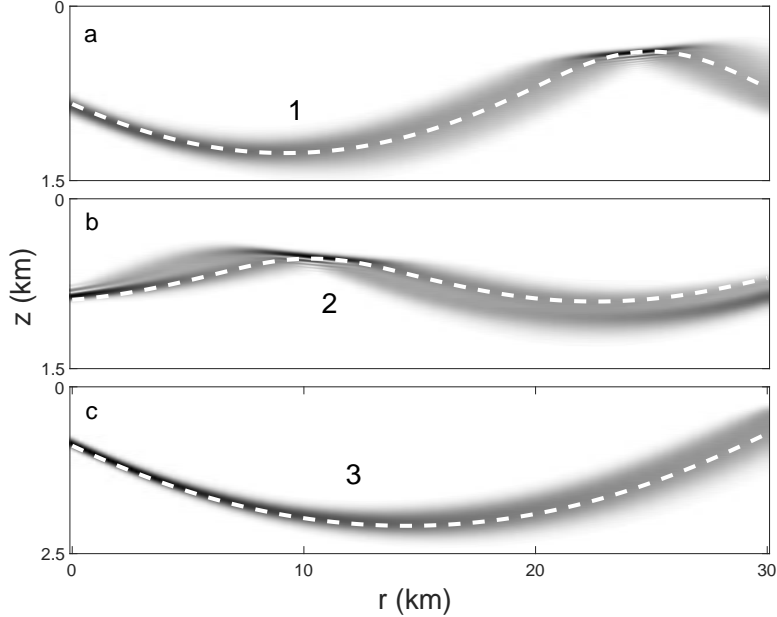}
	\caption{\label{fig:FIG11} The wave beams propagating along the paths shown by the dashed lines. The paths represent the central rays of the 1-st (a), 2-nd (b), and 3-rd (c) eigenbeams.
	}
\end{figure*}

Our idea is to use as the initial fields the components $u_{n}^{\ast}\left(
z\right)  $ corresponding to individual eigenbeams. The antenna with an
initial field $u_{n}^{\ast}\left(  z\right)  $ should emit a wave beam
propagating along the $n$-th eigenbeam. Figures 11a, b, and c show the wave
fields excited by our antenna with $u_{0}\left(  z\right)  $ equal to
$u_{1}^{\ast}\left(  z\right)  B\left(  z\right)  $, $u_{2}^{\ast}\left(
z\right)  B\left(  z\right)  $, and $u_{3}^{\ast}\left(  z\right)  B\left(
z\right)  $, respectively, where $u_{1}\left(  z\right)  $, $u_{2}\left(
z\right)  $, and $u_{3}\left(  z\right)  \,$\ are the field components whose
real parts are shown in Fig. 3. These beams are calculated in the unperturbed
waveguide. In the presence of the sound speed fluctuations, they change
insignificantly (not shown).

The simulation results presented in Fig. 11 show that the proposed method
makes it possible to excite narrow wave beams associated with individual
eigenbeams, and thereby use the antenna in the inhomogeneous medium as an
acoustic searchlight.

\section{Conslusion \label{sec:conclusion}}

The problem of beamforming in an inhomogeneous environment in this paper is addressed using the notion of the stable components of the sound field introduced in Ref. \cite{V2017}. The assumptions underlying this notion do not yet have a rigorous justification and are based only on simple estimates derived in the geometrical optics approximation. Nevertheless, the numerical simulation confirms that the components, called stable, are indeed less sensitive to the sound speed variations than the total wave field \cite{V2017a,V2018,V2019a}.

In Sec. \ref{sec:MFP} it is shown that the use of the field representation as a superposition of the stable components makes it possible to modify the traditional MFP method and make it less sensitive to the inevitable inaccuracies of the environmental model. This approach can be applied to develop robust methods of source localization and reconstruction of environmental parameters. In Refs. \cite{V2017a} and \cite{V2019a}, the stable components have already been used for solving similar problems. However, the approaches considered in these works imply the coherent state expansion of not only the calculated fields but the measured field as well. This requires a long antenna, whose size significantly exceeds the scale of the coherent state $\Delta_{z}$. In contrast, in the present paper, only the calculated fields are decomposed into the coherent states. The vectors $\mathbf{u}_{n}$ are determined by fragments of the calculated stable components within the depth interval overlapped by the antenna. However, in this case, there are still restrictions on the antenna length. On a very short antenna, the vectors $\mathbf{u}_{n}$ with different $n$ become indistinguishable. To fulfill the condition (\ref{unm}), the antenna aperture should be large enough.

The isolation of stable components from the total field excited by a pulsed source is considered in Sec. \ref{sec:pulse}. Here, the pulses $G_{n}\left(t\right)  $ are defined, which are interpreted as signals arriving at the antenna through individual eigenbeams. In the presence of perturbation, their arrival times acquire different delays that carry information about the inhomogeneities through which the eigenbeams pass.

The main attention in this paper is given to the beamforming for a receiving antenna. But in Sec. \ref{sec:emit} it is demonstrated numerically that the use of the stable components to form the amplitude-phase distributions on the antenna elements allows one to emit narrow wave beams propagating along given ray paths.

	The author is grateful to Dr. L. Ya. Lyubavin and Dr. A.
	Yu. Kazarova for valuable discussions.
	The research was carried out within the state assignment of IAP RAS (Project 0035-2019-0019).

\end{document}